\documentclass[twocolumn,aps,pra,floatfix]{revtex4-2}
\usepackage{graphicx}
\usepackage{times}
\usepackage{nicefrac}
\usepackage{amsmath}
\usepackage{amsfonts}
\usepackage{amssymb}
\usepackage{amsthm}
\usepackage{epsf}
\usepackage{bm}
\usepackage{bbm}
\usepackage{times}
\usepackage{physics}

\usepackage{dcolumn}

\newcolumntype{.}{D{x}{}{-1}}
\newcolumntype{w}[1]{D{.}{.}{#1}}

\begin{document}

\newcommand{\half}{\frac12}
\newcommand{\vare}{\varepsilon}
\newcommand{\eps}{\epsilon}
\newcommand{\pr}{^{\prime}}
\newcommand{\ppr}{^{\prime\prime}}
\newcommand{\pp}{{p^{\prime}}}
\newcommand{\ppp}{{p^{\prime\prime}}}
\newcommand{\hp}{\hat{\bfp}}
\newcommand{\hr}{\hat{\bfr}}
\newcommand{\hk}{\hat{\bfk}}
\newcommand{\hx}{\hat{\bfx}}
\newcommand{\hpp}{\hat{\bfpp}}
\newcommand{\hq}{\hat{\bfq}}
\newcommand{\rqq}{{\rm q}}
\newcommand{\bfk}{{\bm{k}}}
\newcommand{\bfp}{{\bm{p}}}
\newcommand{\bfq}{{\bm{q}}}
\newcommand{\bfr}{{\bm{r}}}
\newcommand{\bfx}{{\bm{x}}}
\newcommand{\bfy}{{\bm{y}}}
\newcommand{\bfz}{{\bm{z}}}
\newcommand{\bfpp}{{\bm{\pp}}}
\newcommand{\bfppp}{{\bm{\ppp}}}
\newcommand{\balpha}{\bm{\alpha}}
\newcommand{\bvare}{\bm{\vare}}
\newcommand{\bgamma}{\bm{\gamma}}
\newcommand{\bGamma}{\bm{\Gamma}}
\newcommand{\bLambda}{\bm{\Lambda}}
\newcommand{\bmu}{\bm{\mu}}
\newcommand{\bnabla}{\bm{\nabla}}
\newcommand{\bvarrho}{\bm{\varrho}}
\newcommand{\bsigma}{\bm{\sigma}}
\newcommand{\bTheta}{\bm{\Theta}}
\newcommand{\bphi}{\bm{\phi}}
\newcommand{\bomega}{\bm{\omega}}
\newcommand{\intzo}{\int_0^1}
\newcommand{\intinf}{\int^{\infty}_{-\infty}}
\newcommand{\lbr}{\langle}
\newcommand{\rbr}{\rangle}
\newcommand{\ThreeJ}[6]{
        \left(
        \begin{array}{ccc}
        #1  & #2  & #3 \\
        #4  & #5  & #6 \\
        \end{array}
        \right)
        }
\newcommand{\SixJ}[6]{
        \left\{
        \begin{array}{ccc}
        #1  & #2  & #3 \\
        #4  & #5  & #6 \\
        \end{array}
        \right\}
        }
\newcommand{\NineJ}[9]{
        \left\{
        \begin{array}{ccc}
        #1  & #2  & #3 \\
        #4  & #5  & #6 \\
        #7  & #8  & #9 \\
        \end{array}
        \right\}
        }
\newcommand{\Vector}[2]{
        \left(
        \begin{array}{c}
        #1     \\
        #2     \\
        \end{array}
        \right)
        }

\newcommand{\Dmatrix}[4]{
        \left(
        \begin{array}{cc}
        #1  & #2   \\
        #3  & #4   \\
        \end{array}
        \right)
        }
\newcommand{\Dcase}[4]{
        \left\{
        \begin{array}{cl}
        #1  & #2   \\
        #3  & #4   \\
        \end{array}
        \right.
        }

\newcommand{\Za}{{Z \alpha}}
\newcommand{\im}{{ i}}

\title{QED $\bm{m\alpha^7}$ effects for triplet states of helium-like ions}

\author{Vladimir A. Yerokhin}
\affiliation{Peter the Great St.~Petersburg Polytechnic University,
Polytekhnicheskaya 29, 195251 St.~Petersburg, Russia}

\author{Vojt\v{e}ch Patk\'o\v{s}}
\affiliation{Faculty of Mathematics and Physics, Charles University,  Ke Karlovu 3, 121 16 Prague 2, Czech Republic}

\author{Krzysztof Pachucki}
\affiliation{Faculty of Physics, University of Warsaw,
             Pasteura 5, 02-093 Warsaw, Poland}


\begin{abstract}

We perform {\em ab initio} calculations of the QED effects of order $m\alpha^7$ for the $2^3S$ and $2^3P$ states of
He-like ions. The computed effects are combined with previously calculated energies
from [V.~A.~Yerokhin and K.~Pachucki, Phys.~Rev.~A {\bf 81},
022507 (2010)], thus improving the theoretical accuracy by an order of magnitude.
The obtained theoretical values for the $2\,^3S$-$2\,^3P_{0,2}$ transition energies are
in good agreement with available experimental results and with previous
calculations performed to all orders in the nuclear binding strength parameter $\Za$.
For the ionization energies, however, we find some inconsistency between the
$\Za$-expansion and all-order calculations, which might be related to a similar
discrepancy between the theoretical and experimental results
for the ionization energies of helium [V.~Patk\'o\v{s} {\em et al.},
Phys.~Rev.~A {\bf 103}, 042809 (2021)].

\end{abstract}

\maketitle

\section{Introduction}

Significant progress has recently been achieved in the theoretical description of the Lamb shift in
the helium atom. After extensive efforts, a complete calculation of the QED effects of order $m\alpha^7$
has been accomplished for the triplet states of the helium atom 
\cite{yerokhin:18:betherel,patkos:20,patkos:21,patkos:21:helamb}.
This calculation improved the accuracy of the theoretical energies
of the $2\,^3S$ and $2\,^3P$ states of helium by more than an order of magnitude and made the theoretical
predictions sensitive to the nuclear charge radius on the 1\% level.
The theoretical result for the $2\,^3S$-$2\,^3P$ transition energy was found to be in excellent
agreement with the experimental value \cite{zheng:17}. However, the individual ionization energies of the
$2\,^3S$ and $2\,^3P$ states were shown to deviate by 10$\sigma$ from the experimental results
\cite{clausen:21}.

In the present work we extend our calculations of the $m\alpha^7$ effects from helium to helium-like ions.
The goal of this investigation is twofold. First, our calculations will improve the theoretical accuracy of
the $2\,^3S$-$2\,^3P$ transition energies in light He-like ions. This is of particular importance in the case
of Li$^+$, for which very precise experimental results are available \cite{riis:94}. Second,
calculations of the $m\alpha^7$ effects for different nuclear charges $Z$ will allow us to study
the $Z$-dependence of this correction (in particular, the high-$Z$ asymptotics) and to perform a
cross-check against the hydrogen theory and independent calculations carried out to all orders 
in the nuclear binding strength parameter $\Za$.

\section{General formulas}

The QED effects of order $m\alpha^7$ for the centroid energy of triplet states of helium-like atoms
were derived by us in a series of works \cite{yerokhin:18:betherel,patkos:20,patkos:21,patkos:21:helamb}.
In this paper we transform the obtained formulas to a form that is relatively compact and more suitable for
studying the $Z$-dependence of these effects.

Formulas derived in previous works contained logarithmic contributions of two types, specifically,
$\ln(\Za)$ in the electron-nucleus terms and $\ln(\alpha)$ in the electron-electron terms. In addition,
there were terms with $\ln(Z)$ implicitly present in matrix elements of individual operators and
the Bethe-logarithm contributions.
In the present work we show that the complete dependence of the $m\alpha^7$ correction
on $\ln(Z)$ and $\ln(\alpha)$ can be
factorized out in terms of $\ln(\Za)$ and $\ln^2(\Za)$. The exact matching of coefficients at $\ln(Z)$ and $\ln(\alpha)$
in the electron-electron terms served as an important cross-check of our derivation.

The QED correction of order $m\alpha^7$ for the centroid energy of triplet states of helium-like atoms
is represented as a sum of the double-logarithmic, single-logarithmic, and non-logarithmic contributions,
\begin{eqnarray}\label{eq:1}
E^{(7)} = E^{(7,2)}\,\ln^2(\Za)^{-2} + E^{(7,1)}\,\ln(\Za)^{-2} + E^{(7,0)}\,,
\end{eqnarray}
where contributions $E^{(7,i)}$ do not contain any logarithms in their $1/Z$ expansion and are defined
as follows,
\begin{align}
E^{(7,2)} = -\frac1{2\pi} Z^3\,Q_1 = -2 Z^3\, \lbr \delta^3(r_1)\rbr\,,
\end{align}
\begin{widetext}
\begin{align}\label{eq:3}
E^{(7,1)} = &\ \frac1{3 \pi} \Bigg[
-8 E_0 E_4 
- \frac{Z}{5}\Big(\frac{19}{3} + 11 Z \Big) Q_{3}
+ \frac{11 Z}{10} Q_{4}
- \frac{39}{10} Q_{6T} +
  4 E_4 Q_{7}
 \nonumber \\ &
  + Z\Big(-\frac{E_0}{5} + \frac{9 Z^2}{8} + 8 Z^2 \ln 2 + Q_{7} \Big) Q_{1}
  +
  \frac{26}{5} Q_{10} + 4 E_0 Z^2 Q_{11} + 8 E_0 Z^2 Q_{12} - 8 E_0 Z Q_{13}
 \nonumber \\ &
  -  8 Z^2 Q_{14}
  + 8 Z^3 Q_{15} - 4 Z^2 Q_{16} + 4 Z Q_{17} - \frac{38 Z}{5} Q_{18}
   +
  2 Z^2 Q_{21} + 2 Z^2 Q_{22} + 4 Z Q_{24} - 2 Z Q_{28}
 \nonumber \\ &
  + \frac{11 Z}{10} Q_{51} +
	     4 E_0^2 Z Q_{53} - \frac{Z}{5} Q_{62} + 3 Z^2 \widetilde{Q}_{57}
+ 2\, \bigg\langle H_R'\frac{1}{(E_0-H_0)'}H_R\bigg\rangle
	     \Bigg] \,,
\end{align}
\begin{align}\label{eq:4}
E^{(7,0)} = &\
 \frac1{90 \pi}
\Bigg\{
-8 E_0 E_4 \big( 19 - 30 \ln 2\big) 
+ Z\Big(-\frac{53183}{420} - \frac{2003 Z}{140} +  82 \ln 2 + 66 Z \ln 2 \Big) Q_{3}
 \nonumber \\ &
    + Z\Big(\frac{2003}{280} - 33 \ln 2 \Big) Q_{4}
    +  \Big(\frac{14971}{70} + 36 \ln 2\Big) Q_{6T}
  + \Big(76 E_4 - 120 E_4 \ln 2 - 105 Q_{9}\Big) Q_{7}
 \nonumber \\ &
  +
  \Big(\frac{9543}{20} - 264 \ln 2\Big) Q_{10} +
  4 E_0 Z^2 \Big(19  - 30 \ln 2\Big) Q_{11} +
  8 E_0 Z^2\Big(19  - 30 \ln 2\Big) Q_{12}
 \nonumber \\ &
  -8 E_0 Z \Big(19  - 30 \ln 2\Big)  Q_{13} 
  -8 Z^2 \Big(19 - 30 \ln 2\Big) Q_{14} +
   8 Z^3 \Big(19 - 30 \ln 2\Big)  Q_{15}
 \nonumber \\ &
  - 4 Z^2 \Big(19  - 30 \ln 2\Big) Q_{16} +
   4 Z  \Big(19 - 30 \ln 2\Big) Q_{17} +
  Z \Big(-\frac{2757}{10} + 288 \ln 2 \Big) Q_{18}
 \nonumber \\ &
  + 2 Z^2 \Big(19 - 30  \ln 2\Big) Q_{21} +
  2 Z^2 \Big(19  - 30  \ln 2\Big) Q_{22} +
  4 Z \Big(19 - 30 \ln 2\Big) Q_{24} +
  \frac{105}{8} Q_{25}
 \nonumber \\ &
   -  2 Z \Big(19 - 30 \ln 2\Big) Q_{28}
  +
  Z \Big(\frac{3893}{280} - 33 \ln 2 \Big) Q_{51} +
  Z \Big(76 E_0^2 - 120 E_0^2 \ln 2 + 105 Q_{9}\Big)  Q_{53}
 \nonumber \\ &
   - 105 Z Q_{59}
   +
  \frac{105}{4} Q_{61} +
  4 Z \Big(\frac{7}{5} + 3 \ln 2\Big) Q_{62} +
  88 Z \widetilde{Q}_{52} - 72 \widetilde{Q}_{54} - 297 \widetilde{Q}_{55}
 \nonumber \\ &
   +
  Z^2 \Big( \frac{513}{4} - 90 \ln 2 \Big) \widetilde{Q}_{57}
  - 24 Z \widetilde{Q}_{58} - 63 \widetilde{Q}_{60} +
  12 Z \widetilde{Q}_{63}
  + 
  Z \Bigg[
  \frac{3317 E_0}{140} + \frac{5755 Z^2}{56}
    \nonumber \\ &
- \frac{85 \pi^2 Z^2}{6} +
    6 E_0 \ln 2 - 362\, Z^2 \ln 2 + 45\, Z^2 \ln^2 2 +
    (19 - 30 \ln 2) Q_{7} + \frac{225 Z^2}{2} \zeta(3) \Bigg]\, Q_{1}
    \Bigg\}
    \nonumber \\ &
 + \frac{Z^3}{2\pi}\beta_L\, Q_{1} + \frac{Z^2}{2\pi^2}\, B_{50}\,Q_{1} + \frac{Z}{2\pi^3}\, C_{40}\,Q_{1}
 + E_{\rm sec}\,.
\end{align}
\end{widetext}
In the above formulas, $Q_1\ldots Q_{64}$ are the expectation values of the basic elementary operators defined
in Table~\ref{oprsQ}. Some of $Q_i$ contain implicitly terms with $\ln(Z)$, which need to be separated out.
We thus introduced expectation values $\widetilde{Q}_{i}$, which are free from $\ln(Z)$ and are defined
by
\begin{align}
{Q}_{52} &\ = \widetilde{Q}_{52} + \frac12 \ln Z^{-2} \, Q_{3}\,,\\
{Q}_{54} &\ = \widetilde{Q}_{54} + \frac12 \ln Z^{-2} \, Q_{10}\,,\\
{Q}_{55} &\ = \widetilde{Q}_{55} + \frac16 \ln Z^{-2} \, Q_{6T}\,,\\
{Q}_{56} &\ = \widetilde{Q}_{56} + \frac12 \ln Z^{-2} \, Q_{1}\,,
\end{align}
\begin{align}
{Q}_{57} &\ = \widetilde{Q}_{57} - Z \ln Z^{-2} \, Q_{1}\,,\\
{Q}_{58} &\ = \widetilde{Q}_{58} + \frac12 \ln Z^{-2} \, Q_{18}\,,\\
{Q}_{60} &\ = \widetilde{Q}_{60} + \frac12 \ln Z^{-2} \, Q_{6T}\,,\\
{Q}_{63} &\ = \widetilde{Q}_{63} + \frac12 \ln Z^{-2} \, Q_{62}\,.
\end{align}
Further notations in Eqs.~(\ref{eq:3}) and (\ref{eq:4}) are as follows:
$E_0$ is the nonrelativistic energy, $E_4$ is the leading relativistic (Breit) correction of order $m\alpha^4$,
$\beta_L$ is the relativistic Bethe-logarithm correction defined as in Ref.~\cite{yerokhin:22:bethe},
$B_{50} = -21.554\,47$ and $C_{40} = 0.417\,503\,770$ are the hydrogenic
two-loop $(\Za)^5$ and three-loop $(\Za)^4$ expansion coefficients, respectively, see Ref.~\cite{yerokhin:18:hydr},
and $E_{\rm sec}$ is the second-order correction given by
\begin{align}
E_{\rm sec} = &\ 2\,\Big< H_{\rm fs}^{(5)}\frac{1}{(E_0-H_0)'}H_{\rm fs}^{(4)}\Big>
 \nonumber \\ &
 + \frac1{\pi}\Big( \frac{19}{45} - \frac23 \ln 2\Big)
 \Big< H_R'\frac{1}{(E_0-H_0)'}H_R\Big>
 \nonumber \\ &
 -\frac{7}{3\pi}\Big< \frac1{r^3} \frac{1}{(E_0-H_0)'}H_R\Big>\,.
\end{align}
The effective Hamiltonians in the above formulas are defined as follows. $H_R$ is a regular
part of the spin-independent Breit
Hamiltonian and is defined by its action on a ket eigenstate  $|\phi\rbr$ of the nonrelativistic Hamiltonian with
the energy $E$ as
\begin{align}
H_R|\phi\rangle &\ =\bigg[-\frac12(E-V)^2-\frac{Z}{4}\frac{\vec r_1\cdot\vec\nabla_1}{r_1^3}
-\frac{Z}{4}\frac{\vec r_2\cdot\vec\nabla_2}{r_2^3}
 \nonumber\\&
 +\frac14\nabla_1^2\nabla_2^2
+\nabla_1^i\frac{1}{2r}\bigg(\delta^{ij}+\frac{r^i r^j}{r^2}\bigg)\,\nabla_2^j\bigg]|\phi\rangle\,, \label{13}
\end{align}
where $V= - Z/r_1 - Z/r_2 + 1/r$. The operator $H_R'$ is defined by its action on a ket state $|\phi\rbr$ as
\begin{align}
H_R'|\phi\rangle = -2Z\bigg(\frac{\vec r_1\cdot\vec\nabla_1}{r_1^3}+\frac{\vec r_2\cdot\vec\nabla_2}{r_2^3}\bigg)|\phi\rangle\,.
\end{align}
The operators $H_{\rm fs}^{(4)}$ and $H_{\rm fs}^{(5)}$ are the $m\alpha^4$ and
$m\alpha^5$ parts of the spin-dependent Breit Hamiltonian $H_{\rm fs}$ with anomalous magnetic moment,
correspondingly,
\begin{align}
H_{\rm fs} = \alpha^4 H_{\rm fs}^{(4)} + \alpha^5 H_{\rm fs}^{(5)} + O(\alpha^6)\,,
\end{align}
\begin{eqnarray} \label{fs}
H_{\rm fs} & = &\frac{\alpha}{4\,m^2}\left(
\frac{\vec{\sigma}_1\cdot\vec{\sigma}_2}{r^3}
-3\,\frac{\vec{\sigma}_1\cdot\vec{r}\,
\vec{\sigma}_2\cdot\vec{r}}{r^5}\right)(1+\kappa)^2\, \nonumber \\
& + & {Z\alpha  \over 4 m^2}
\left[
\frac{1}{r_1^3}\,\vec{r}_1\times\vec{p}_1\cdot\vec{\sigma}_1+
\frac{1}{r_2^3}\,\vec{r}_2\times\vec{p}_2\cdot\vec{\sigma}_2
\right](1+2\kappa)
\nonumber \\
& + &
\frac{\alpha}{4\,m^2
\,r^3}\biggl[
\bigl[(1+2\,\kappa)\,\vec{\sigma}_2+2\,(1+\kappa)\,\vec{\sigma}_1\bigr]\cdot\vec{r}\times\vec{p}_2
\nonumber \\ &&
-\bigl[(1+2\,\kappa)\,\vec{\sigma}_1+2\,(1+\kappa)\,\vec{\sigma}_2\bigr]\cdot\vec{r}
\times\vec{p}_1\biggr]\,,
\end{eqnarray}
where $\kappa = \alpha/(2\pi)+O(\alpha^2)$ is the anomalous magnetic moment of the electron.

\begin{table*}
\caption{Definitions of elementary basic operators $Q_i$. Notations are: $r \equiv |\vec{r}_1-\vec{r}_2|$,
$\vec P = \vec{p}_1+\vec{p}_2$, $\vec p = \nicefrac12
\big(\vec{p}_1-\vec{p}_2\big)$.}
\label{oprsQ}
\begin{ruledtabular}
\begin{tabular}{llll}
$Q_1 $ & $4 \pi \delta^3 (r_1)$   				     &$Q_{33}$ & $\vec{p}_1\cdot\vec{p}_2$			     \\
$Q_2 $ & $4 \pi \delta^3 (r)$               	     &$Q_{34}$ & $\vec{P}\,/r_1\,\vec{P}$				 \\
$Q_3 $ & $4 \pi \delta^3(r_1)/r_2$                   &$Q_{35}$ & $\vec{P}\,/r\,\vec{P}$				     \\
$Q_4 $ & $4 \pi \delta^3(r_1)\, p_2^2$ 	             &$Q_{36}$ & $\vec{P}\,/r_1^2\,\vec{P}$              \\
$Q_5 $ & $4 \pi \delta^3(r)/r_1$				     &$Q_{37}$ & $\vec{P}\,/(r_1 r_2)\,\vec{P}$			 \\
$Q_{6T} $ & $4 \pi\,\vec{p}\,\delta^3(r)\,\vec{p} $	 &$Q_{38}$ & $\vec{P}\,/(r_1 r)\,\vec{P}$			 \\
$Q_7 $ & $1/r$						                 &$Q_{39}$ & $\vec{P}\,/r^2\,\vec{P}$				    \\
$Q_8 $ & $1/r^2$						             &$Q_{40}$ & $p_1^2\,p_2^2\,P^2$				         \\
$Q_9 $ & $1/r^3$                    	             &$Q_{41}$ & $P^2\,p_1^i\, (r^i r^j + \delta^{ij} r^2)/r^3 \, p_2^j$  \\
$Q_{10}$ & $1/r^4$                  	             &$Q_{42}$ & $p_1^i\,(r_1^i r_1^j + \delta^{ij} r_1^2)/r_1^4 \,  P^j$ \\
$Q_{11}$ & $1/r_1^2$                	             &$Q_{43}$ & $p_1^i\,(r_1^i r_1^j + \delta^{ij} r_1^2)/(r_1^3 r_2)\,  P^j$ \\
$Q_{12}$ & $1/(r_1 r_2)$            	             &$Q_{44}$ & $p_1^i\,p_2^k\,(r_1^ir_1^j+\delta^{ij}r_1^2)/r_1^3\,p_2^k\, P^j$\\
$Q_{13}$ & $1/(r_1 r)$              	             &$Q_{45}$ & $p_2^i(r^i r^j+\delta^{ij} r^2)(r_1^jr_1^k+\delta^{jk} r_1^2)/(r_1^3 r^3)\, P^k$	\\
$Q_{14}$ & $1/(r_1 r_2 r)$          	             &$Q_{46}$ & $p_1^i(r_1^i r_1^j+\delta^{ij} r_1^2)(r_2^jr_2^k+\delta^{jk} r_2^2)/(r_1^3 r_2^3)\, p_2^k$\\
$Q_{15}$ & $1/(r_1^2 r_2)$					         &$Q_{47}$ & $(\vec{r}_1\cdot\vec{r}_2)/(r_1^3 r_2^2)$             \\
$Q_{16}$ & $1/(r_1^2 r)$					         &$Q_{48}$ & $r_1^i r^j(r_1^i r_1^j-3\delta^{ij} r_1^2)/(r_1^4r^3)$\\
$Q_{17}$ & $1/(r_1 r^2)$   					         &$Q_{49}$ & $r_1^ir^j(r_2^ir_2^j-3\delta^{ij}r_2^2)/(r_1^3r_2r^3)$\\
$Q_{18}$ & $(\vec{r}_1\cdot\vec r)/(r_1^3 r^3)$      &$Q_{50}$ & $p_2^k\,r_1^i/r_1^3\,(\delta^{jk}r_2^i/r_2-\delta^{ik}r_2^j/r_2-\delta^{ij}r_2^k/r_2-r_2^ir_2^jr_2^k/r_2^3)\,p_2^j$\\
$Q_{19}$ & $(\vec{r}_1\cdot\vec r)/(r_1^3 r^2)$      &$Q_{51} $ & $4 \pi\,\vec p_1\,\delta^3(r_1)\,\vec p_1 $	   \\
$Q_{20}$ & $r_1^i r_2^j(r^i r^j-3\delta^{ij}r^2)/(r_1^3 r_2^3 r)$
                                                     &$Q_{52} $ & $4 \pi \delta^3(r_1)/r_2\,(\ln r_2+\gamma)$      \\
$Q_{21}$ & $p_2^2/r_1^2$					         &$Q_{53} $ & $1/r_1$						                    \\
$Q_{22}$ & $\vec{p}_1/r_1^2\, \vec{p}_1$			 &$Q_{54}$ & $1/r^4 (\ln r+\gamma)$                  	         \\
$Q_{23}$ & $\vec{p}_1/r^2\, \vec{p}_1$			     &$Q_{55}$ & $1/r^5$                  	                     \\
$Q_{24}$ & $p_1^i\,(r^i r^j+\delta^{ij} r^2)/(r_1 r^3)\, p_2^j$
                                                     &$Q_{56}$ & $1/r_1^3$   					                    \\
$Q_{25}$ & $P^i\, (3 r^i r^j-\delta^{ij} r^2)/r^5\, P^j$	
                                                     &$Q_{57}$ & $1/r_1^4$   					                     \\
$Q_{26}$ & $p_2^k \,r_1^i\,/r_1^3 (\delta^{jk} r^i/r - \delta^{ik} r^j/r -\delta^{ij} r^k/r -r^i
r^j r^k/r^3)\, p_2^j$		                         &$Q_{58}$ & $(\vec{r}_1\cdot\vec r)/(r_1^3 r^3)(\ln r+\gamma)$ \\
$Q_{27}$ & $p_1^2\, p_2^2$					         &$Q_{59}$ & $1/(r_1 r^3)$				                      \\
$Q_{28}$ & $p_1^2\,/r_1\, p_2^2$				     &$Q_{60}$ & $\vec p\,/r^3\, \vec p$				             \\
$Q_{29}$ & $\vec{p}_1\times\vec{p}_2\,/r\,\vec{p}_1\times\vec{p}_2$
                                                     &$Q_{61}$ & $\vec P\,/r^3\, \vec P$				              \\
$Q_{30}$ & $p_1^k \,p_2^l\,(-\delta^{jl} r^i r^k/r^3 - \delta^{ik} r^j r^l/r^3 +3r^i r^j r^k
r^l/r^5)\, p_1^i\, p_2^j$			                 &$Q_{62}$ & $r^i r^j(\delta^{ij}r_1^2-3r_1^i r_1^j)/(r_1^5 r^3)$	\\
$Q_{31}$ & $4 \pi \delta^3(r_1)\, \vec{p}_1\cdot\vec{p}_2$
                                                     &$Q_{63}$ & $r^i r^j(\delta^{ij}r_1^2-3r_1^i r_1^j)/(r_1^5 r^3) (\ln r+\gamma)$				 \\
$Q_{32}$ & $(\vec{r}_1\cdot\vec{r}_2)/(r_1^3 r_2^3)$ &$Q_{64}$ & $p^i (\delta^{ij}r^2-3r^i r^j)/r^5 p^j$              \\
\end{tabular}
\end{ruledtabular}
\end{table*}

\section{Higher-order effects}
\label{sec:ho}

The effects of order $m\alpha^8$ and higher cannot be calculated rigorously at present and
need to be estimated. Our approximation for these effects is represented
as a sum of three terms,
\begin{align}
E^{(8+)} = E^{(8)}_D + E^{(8)}_{\rm 1ph} + E^{(8+)}_{\rm rad}\,,
\end{align}
where $E^{(8+)}_D$ comes from the one-electron Dirac energy, $E^{(8+)}_{\rm 1ph}$
originates from the one-photon exchange correction, and $E^{(8+)}_{\rm rad}$ represents
the radiative QED effects.

The Dirac contribution to the ionization energy of an $1snl$ state comes from the valence
electron, $E_D = E_D(nl)$ and is given by
\begin{align}
E^{(8)}_D(2s) =E^{(8)}_D(2p_{1/2}) = -\frac{429}{32768}\,Z^8\,, \\
E^{(8)}_D(2p_{3/2}) = -\frac{5}{32768}\,Z^8\,.
\end{align}

The one-photon exchange correction of order $m\alpha^8$ was calculated in Ref.~\cite{mohr:85:pra},
with the result
\begin{align}
E^{(8)}_{\rm 1ph}\left(2^3S\right) =   &\ 0.0281\,Z^7\,, \\
E^{(8)}_{\rm 1ph}\left(2^3P_0\right) = &\ 0.1070\,Z^7\,, \\
E^{(8)}_{\rm 1ph}\left(2^3P_2\right) = &\ 0.0037\,Z^7\,.
\end{align}
We note a relative large numerical contribution of the one-photon exchange correction for the $2^3P_0$
state.

An approximation for the radiative QED contribution of order $m\alpha^8$ and higher is
obtained by scaling the hydrogenic results with the expectation value of the
$\delta$-function \cite{drake:88:cjp,yerokhin:10:helike},
\begin{align}\label{eq:hydr:appr}
E^{(8+)}_{\rm rad} = &\ \Big[ E^{(8+)}_{\rm rad, H}(1s) + E^{(8+)}_{\rm rad, H}(nl)\Big]\,
 \frac{\lbr \sum_i\delta^3(r_i) \rbr}{\frac{Z^3}{\pi}\left( 1 + \frac{\delta_{l,0}}{n^3}\right)}
\nonumber \\ &
 -  E^{(8+)}_{\rm rad, H}(1s)\,,
\end{align}
where $E^{(8+)}_{\rm rad, H}(nl)$ is the hydrogenic QED contribution of order order $m\alpha^8$
and higher of an $nl$ state. This contribution consists of the one-loop and two-loop effects,
which are reviewed in Ref.~\cite{yerokhin:18:hydr}.
We estimate the uncertainty of this approximation for He-like ions as 75\% of the few-body part
of $E^{(8+)}_{\rm rad}$, specifically,
\begin{align}
\delta E^{(8+)}_{\rm rad} =
\pm 0.75\, \Big[ E^{(8+)}_{\rm rad, H}(1s) &\, +  E^{(8+)}_{\rm rad, H}(nl)\Big]\,
 \nonumber \\ & \times
 \Bigg[ \frac{\lbr \sum_i\delta^3(r_i) \rbr}{\frac{Z^3}{\pi}\left( 1 + \frac{\delta_{l,0}}{n^3}\right)}
 -1\Bigg]\,.
\end{align}

In addition we include the finite nuclear size correction, which is obtained from the
corresponding hydrogenic corrections analogously to Eq.~(\ref{eq:hydr:appr}),
see Ref.~\cite{yerokhin:10:helike} for details.

%
\section{Numerical results}

In this work we performed calculations of the $m\alpha^7$ effects for the centroid energies of the $2{}^3S$ and $2{}^3P$
states of helium-like ions with $Z \le 12$.
The computation followed the numerical approach developed in our previous investigations \cite{yerokhin:10:helike,patkos:21:helamb}
and used results for the relativistic Bethe-logarithm correction obtained in Ref.~\cite{yerokhin:22:bethe}.

Numerical values for the $m\alpha^7$ corrections to energies of the
$2{}^3S$, $2{}^3P_0$, and $2{}^3P_2$ states of helium and helium-like ions are presented in Table~\ref{tab:ma7}.
Results for the $2{}^3P_{0,2}$ states are obtained by combining the $m\alpha^7$ correction for the $2{}^3P$ centroid
energy calculated in this work and the corresponding corrections to the fine structure from
Ref.~\cite{pachucki:10:hefs}. We do not present results for the $2{}^3P_{1}$ state because it mixes with the
$2{}^1P_{1}$ state and thus requires a separate treatment \cite{yerokhin:22:helike}.
Results for helium listed in Table~\ref{tab:ma7} are in full agreement with those reported by us previously
\cite{patkos:21:helamb}.

Table~\ref{tab:ma7} also presents results for the coefficients of the $1/Z$ expansion of the $m\alpha^7$ contributions,
\begin{align}\label{eq:18}
E^{(7,i)} = Z^6 \Big(c^{(7,i)}_0 + \frac{c^{(7,i)}_1}{Z} + \frac{c^{(7,i)}_2}{Z^2} + \ldots \Big)\,.
\end{align}
The leading coefficients $c^{(7,i)}_0$ are known from the hydrogen theory. They are induced by the one-loop QED
correction of order $\alpha(\Za)^6$. Specifically, for the $1snl_j$ state, we have
\begin{align}\label{eq:27}
c_0^{(7,i)} = \frac1{\pi} \Big[ A_{6i}(1s) + \frac{A_{6i}(nl_j)}{n^3}\Big]\,,
\end{align}
where the coefficients $A_{6i}(nl_j)$ are listed in Ref.~\cite{yerokhin:18:hydr}.

We checked that our formulas for $E^{(7,i)} $ are reduced to $Z^6 c^{(7,i)}_0$ in the large-$Z$ limit, see
Appendix~\ref{sec:app} for details. We also checked this correspondence for our numerical results,
by fitting the numerical data from
Table~\ref{tab:ma7} to the form (\ref{eq:18}) and comparing the fitted values of the coefficients $c^{(7,i)}_0$
with the analytical result of Eq.~(\ref{eq:27}).  In this way we confirmed that our calculations of the $m\alpha^7$ effects are
correct to the leading (zeroth) order in $1/Z$.

As a further test, we will compare the next term of the $1/Z$ expansion of $E^{(7)}$
with results
of the all-order (in $\Za$) calculations performed recently in
Ref.~\cite{yerokhin:22:helike}. In that work results were obtained for the higher-order two-electron
QED remainder function that contains contributions of order $m\alpha^{7+}$ and is linear in $1/Z$.
The remainder function $G_{\rm 2elQED}^{(7+)}(\Za) = \delta E^{(7+)}/[m\alpha^2(\Za)^5]$ is defined by Eqs.~(21)-(23)
of Ref.~\cite{yerokhin:22:helike}.
In the limit $\Za\to 0$,
$G_{\rm 2elQED}^{(7+)}(\Za)$ should approach the linear in $1/Z$ part of $E^{(7)}$, if one removes the two-loop part
that is  not included into the all-order calculations.

The linear in $1/Z$ part of $E^{(7)}$ is induced by the coefficients $c_1^{(7,i)}$. The two-loop effects
influence only the nonlogarithmic coefficient $c_1^{(7,0)}$. The corresponding contribution
comes from the hydrogenic correction $\propto \alpha^2(\Za)^5$ and is given by
\begin{align}
c_1^{(7,0)}({\rm 2loop}) = \frac{B_{50}}{\pi^2}\,\Big(1+ \frac{\delta_{l,0}}{n^3}\Big)\,,
\end{align}
where $B_{50} = -21.554\,47$, see Ref.~\cite{yerokhin:22:helike}. It is interesting that the two-loop
part of $c_1$ is much larger than the total values of $c_1$ in Table~\ref{tab:ma7}, which means that
the corresponding one-loop and two-loop contributions largely cancel each other.

The function $G_{\rm 2elQED}^{(7+)}$ was calculated for $Z \ge 10$ in
Ref.~\cite{yerokhin:22:helike}. The extrapolation of the numerical
values towards smaller values of $Z$ is complicated by presence of logarithms. In order to make an extrapolation possible,
we subtract all known logarithms, introducing a new function $G_{\rm nlog}^{(7+)}$
that has a smooth behaviour in the region  $Z\approx 0$,
\begin{align}\label{eq:ma7nlog}
G_{\rm nlog}^{(7+)}(\Za) = &\ G_{\rm 2elQED}^{(7+)}(\Za) - c^{(7,2)}_1\,\ln^2(\Za)^{-2}
        \nonumber \\ &
        - c^{(7,1)}_1\,\ln(\Za)^{-2}
        - c^{(8,1)}_1\,(\Za) \ln(\Za)^{-2}\,.
\end{align}
The logarithmic coefficient in the order $m\alpha^8$ comes from the one-loop self-energy and
vacuum-polarization contribution $\propto\!\alpha(\Za)^6\ln(\Za)$. It is known for hydrogen
\cite{karshenboim:97,mohr:75:prl}. Since it is proportional to the Dirac $\delta$ function,
the result can be immediately generalized to the few-electron case,
\begin{align}
c^{(8,1)}_1 = \Big(\frac{427}{192} - \ln 2\Big)\,\delta_{1}\,,
\end{align}
where $\delta_{1}$ is the $1/Z^1$ coefficient of the $1/Z$ expansion of the matrix element of the Dirac
$\delta$ function, $\delta_{1}(2{}^3S) = -0.211\,484$ and $\delta_{1}(2{}^3P) = -0.085\,951$
\cite{drake:88:cjp}.

In the $Z \to 0$ limit, the function $G_{\rm nlog}^{(7+)}$ should coincide with the $c_1^{(7,0)}$
coefficient from our $m\alpha^7$ calculations, after subtraction of the two-loop part. Specifically,
\begin{align}\label{eq:ma7nlogNRQED}
G_{\rm nlog}^{(7+)}(Z = 0) = c_1^{(7,0)} - c_1^{(7,0)}({\rm 2loop})\,.
\end{align}

\begin{figure*}
\centerline{
\resizebox{0.99\textwidth}{!}{%
  \includegraphics{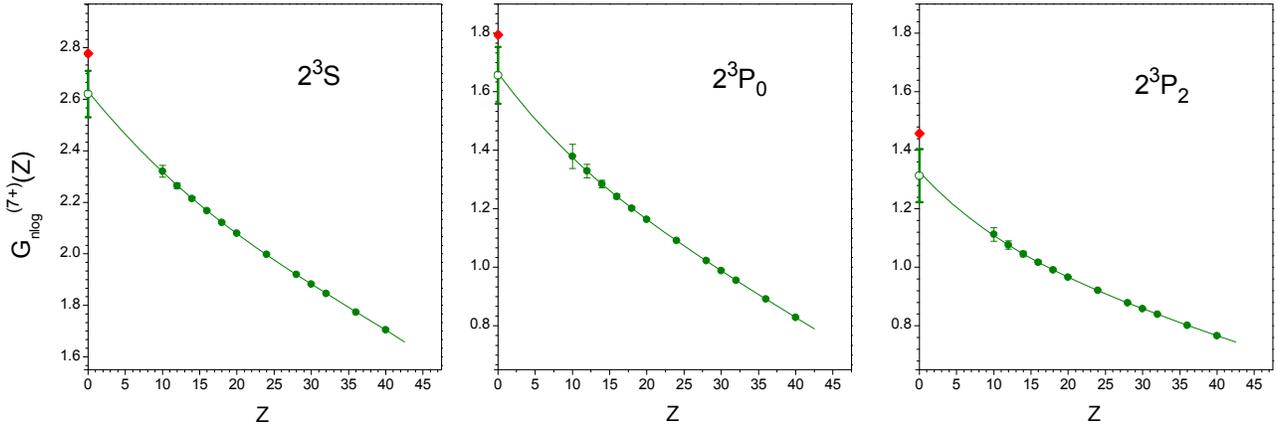}
}
}
 \caption{The nonlogaritmic $m\alpha^{(7+)}$ contribution defined by Eq.~(\ref{eq:ma7nlog}) as a function of the nuclear charge $Z$,
 for the $2^3S$, $2^3P_0$, and $2^3P_2$ states of He-like ions.
 Filled green dots denote results of all-order numerical calculations, open green dots show fitting results at $Z = 0$, red diamonds
 display the $\alpha$-expansion results.
\label{fig:ma7}}
\end{figure*}

In Fig.~\ref{fig:ma7} we present a comparison of numerical values of the function $G_{\rm nlog}^{(7+)}(Z)$
extracted from the all-order calculations of Ref.~\cite{yerokhin:22:helike}
and our present results for the $Z = 0$ limiting value (\ref{eq:ma7nlogNRQED}).
The all-order data were fitted by a polynomial to yield results for the $Z = 0$ limit.
As can be seen from the figure, a small inconsistency between the all-order and
our present $\alpha$-expansion results at $Z = 0$ is observed.
While the deviations are only slightly larger than the estimated uncertainties of the fit,
it is remarkable that for all three states studied
they are of the same sign and of comparable magnitude.
These deviations might be related to the 0.4~MHz difference between the theoretical
and experimental $2^3S$ and $2^3P$ ionization energies of helium
reported in Refs.~\cite{patkos:21:helamb,clausen:21}.
Similarly to the helium case, the deviations
largely cancels in the $2^3S$-$2^3P$ difference.

\begin{table*}
\caption{The $m\alpha^7$ corrections for energies of triplet states of He-like atoms.
\label{tab:ma7}}
\begin{ruledtabular}
\begin{tabular}{lw{3.7}w{3.7}w{5.9}w{5.7}w{3.7}w{5.9}w{3.7}w{5.9}}
\multicolumn{1}{l}{$Z$} &
    \multicolumn{3}{c}{$2{}^3S$} &
        \multicolumn{1}{c}{$2{}^3P$} &
            \multicolumn{2}{c}{$2{}^3P_0$} &
                \multicolumn{2}{c}{$2{}^3P_2$}
        \\\cline{2-4}\cline{5-5}\cline{6-7}\cline{8-9}\\[-5pt]
        & \multicolumn{1}{c}{$ E^{(7,2)}/Z^6$}
            & \multicolumn{1}{c}{$ E^{(7,1)}/Z^6$}
                & \multicolumn{1}{c}{$ E^{(7,0)}/Z^6$}
        & \multicolumn{1}{c}{$ E^{(7,2)}/Z^6$}
            & \multicolumn{1}{c}{$ E^{(7,1)}/Z^6$}
                & \multicolumn{1}{c}{$ E^{(7,0)}/Z^6$}
            & \multicolumn{1}{c}{$ E^{(7,1)}/Z^6$}
                & \multicolumn{1}{c}{$ E^{(7,0)}/Z^6$}
  \\
\hline\\[-5pt]
   2 &   -0.330\,089  &    1.725\,409  &  -11.343\,605\,(7)  &   -0.314\,715  &    1.649\,911  &  -10.825\,73\,(8) &   1.648\,886  &  -10.826\,25\,(8) \\
   3 &   -0.338\,059  &    1.775\,871  &  -11.290\,585\,(7)  &   -0.313\,708  &    1.645\,555  &  -10.458\,59\,(6) &   1.647\,921  &  -10.470\,69\,(6) \\
   4 &   -0.342\,592  &    1.805\,773  &  -11.283\,785\,(7)  &   -0.313\,991  &    1.650\,084  &  -10.314\,51\,(6) &   1.653\,032  &  -10.329\,21\,(6) \\
   5 &   -0.345\,472  &    1.825\,149  &  -11.285\,055\,(7)  &   -0.314\,440  &    1.655\,751  &  -10.242\,46\,(6) &   1.658\,167  &  -10.256\,15\,(6) \\
   6 &   -0.347\,456  &    1.838\,657  &  -11.287\,954\,(7)  &   -0.314\,856  &    1.660\,905  &  -10.200\,67\,(6) &   1.662\,493  &  -10.212\,27\,(6) \\
   7 &   -0.348\,903  &    1.848\,593  &  -11.290\,984\,(7)  &   -0.315\,211  &    1.665\,302  &  -10.173\,90\,(6) &   1.666\,033  &  -10.183\,22\,(6) \\
   8 &   -0.350\,005  &    1.856\,203  &  -11.293\,764\,(7)  &   -0.315\,508  &    1.669\,005  &  -10.155\,55\,(6) &   1.668\,938  &  -10.162\,68\,(6) \\
   9 &   -0.350\,872  &    1.862\,214  &  -11.296\,219\,(11) &   -0.315\,757  &    1.672\,131  &  -10.142\,32\,(6) &   1.671\,348  &  -10.147\,44\,(6) \\
  10 &   -0.351\,571  &    1.867\,082  &  -11.298\,368\,(14) &   -0.315\,967  &    1.674\,790  &  -10.132\,38\,(6) &   1.673\,370  &  -10.135\,70\,(6) \\
  11 &   -0.352\,148  &    1.871\,104  &  -11.300\,241\,(14) &   -0.316\,147   \\
  12 &   -0.352\,630  &    1.874\,482  &  -11.301\,910\,(18) &   -0.316\,302   \\[3pt]
\multicolumn{4}{l}{$1/Z$-expansion coefficients}\\
$c_0$ &
         -0.358\,099  &    1.913\,246     & -11.324\,577     &   -0.318\,310  &    1.705\,367  &  -10.069\,396     &   1.695\,420  &  -10.047\,690 \\
$c_1$ &   0.067\,317  &   -0.482\,89\,(4) &   0.3211\,(4)    &    0.027\,359  &   -0.368\,03\,(5)& -0.3888\,(4)    &  -0.25568\,(5)&  -0.7262\,(10) \\
$c_2$ &  -0.020\,020  &    0.213\,6\,(15) &  -0.562\,(11)    &   -0.038\,518  &    0.6445\,(14) &  -2.4326\,(33)   &   0.3565\,(12)&  -1.4937\,(80)
\end{tabular}
\end{ruledtabular}
\end{table*}

%
\section{Transition energies}

We are now in a position to collect all available theoretical contributions for the transition
energies between the $n = 2$ triplet states in light He-like ions. A systematic calculation of
all QED effects up to order $m\alpha^6$ has been already performed in our previous investigation
\cite{yerokhin:10:helike}. We now add the $m\alpha^7$ correction tabulated in
Table~\ref{tab:ma7} and estimations of higher-order corrections summarized in Sec.~\ref{sec:ho}.

Our theoretical results for the $2\,^3 S$-$2\,^3 P_{0,2}$ transition energies are presented
in Table~\ref{tab:transen}, in comparison with available experimental data and previous
theoretical values. We observe very good agreement with the experimental results for
Li$^+$ \cite{riis:94} and B$^{3+}$ \cite{dinneen:91}, but a significant deviation in the case of Be$^{2+}$
\cite{scholl:93}. It should be noted that the measurement of Ref.~\cite{scholl:93} was already reported
to disagree with theoretical predictions for the fine structure \cite{pachucki:10:hefs}, which calls for
an independent verification of this experiment.

The comparison with our previous calculations of Ref.~\cite{yerokhin:10:helike} shows an excellent
consistency of the results and of the uncertainty estimates. It can be seen that our present calculation
of the $m\alpha^7$ effects improves the theoretical accuracy by an order of magnitude.

It can be seen from Table~\ref{tab:transen}
that for $Z = 5$ our present theoretical values are fully consistent with our recent
results obtained in Ref.~\cite{yerokhin:22:helike}. It is important that
Ref.~\cite{yerokhin:22:helike} utilized a different approach for calculating the effects
of order $m\alpha^7$ and higher. In that work, the higher-order effects were obtained from
the all-order (in $\Za$) calculations, whereas in the present study we calculate the
$m\alpha^7$ effects rigorously with the $\alpha$ expansion and estimate the $m\alpha^{8+}$
effects from the hydrogenic theory. The comparison with results of
Ref.~\cite{yerokhin:22:helike} thus confirms the consistency of two different
approaches for the $2\,^3 S$-$2\,^3 P$ transition energies.

In summary, we reported calculations of the $m\alpha^7$ QED effects for the $2^3S$ and $2^3P$ states of
He-like ions. The $Z$-dependence of the obtained corrections was studied. It was demonstrated
that all terms containing $\ln(Z)$ and $\ln(\alpha)$ in general formulas can be combined together
and expressed in terms of $\ln(\Za)$. The high-$Z$ limit of the calculated $m\alpha^7$ correction
was cross-checked against the analytical results derived from the hydrogen theory.
The linear term of the $1/Z$ expansion of the $m\alpha^7$ correction was
cross-checked against previous calculations performed to all orders in $\Za$. The consistency
of the two approaches was demonstrated for the $2\,^3 S$-$2\,^3 P$ transition energies but
a small deviation was found for the ionization energies.
In the result, we obtain the most accurate theoretical predictions for the $2\,^3S$-$2\,^3P_{0,2}$
transition energies in He-like Li, Be, and B, which are in good agreement with previous theoretical values
and the experimental data for Li and B.

\begin{acknowledgments}
The work was supported by the Russian Science Foundation (Grant No. 20-62-46006).
K.P. and V.P. acknowledge support from the National Science Center (Poland) Grant No. 2017/27/B/ST2/02459.
\end{acknowledgments}

\begin{table}
\caption{Theoretical and experimental $2\,^3 S$-$2\,^3 P$ transition energies, in cm$^{-1}$.
$A$ is the mass number of the isotope.
\label{tab:transen}}
\begin{ruledtabular}
\begin{tabular}{llw{5.10}w{5.10}w{2.8}l}
\multicolumn{1}{l}{$Z$}
& \multicolumn{1}{l}{$A$}
        & \multicolumn{1}{c}{Theory}
        & \multicolumn{1}{c}{Experiment}
        & \multicolumn{1}{c}{Difference}
        & \multicolumn{1}{c}{Ref.}
  \\
\hline\\[-5pt]
\multicolumn{3}{l}{$2\,^3 S_1$--$2\,^3 P_0$} \\[2pt]
%
 3 & 7 & 18\,231.30193\,(10)  & 18\,231.301972\,(14) & -0.00004\,(10) & \cite{riis:94} \\
   &   & 18\,231.3021\,(11)^a   \\[2pt]
 4 & 9 & 26\,864.61052\,(54)  & 26\,864.6120\,(4)    & -0.0015\,(7)   & \cite{scholl:93}  \\
   &   & 26\,864.6114\,(47)^a \\[2pt]
 5 & 11& 35\,393.6244\,(20)   & 35\,393.627\,(13)    & -0.003\,(13)   & \cite{dinneen:91} \\
   &   & 35\,393.6211\,(49)^b \\
   &   & 35\,393.628\,(14)^a\\[5pt]
%
%
%
\multicolumn{3}{l}{$2\,^3 S_1$--$2\,^3 P_2$} \\[2pt]
%
 3 & 7 & 18\,228.19893\,(10) & 18\,228.198963\,(15)& -0.00003\,(10) & \cite{riis:94} \\
   &   & 18\,228.1989\,(10)^a \\[2pt]
 4 & 9 & 26\,867.94512(54)   &  26\,867.9484\,(3) & -0.0033\,(6)    & \cite{scholl:93}  \\
   &   & 26\,867.9450(47)^a \\[2pt]
 5 & 11& 35\,430.0880(20)    &  35\,430.084\,(9)  &  0.004\,(9)     & \cite{dinneen:91} \\
   &   & 35\,430.0876\,(22)^b \\
   &   & 35\,430.088\,(14)^a  \\
%
\end{tabular}
\end{ruledtabular}
$^a$~Yerokhin and Pachucki 2010 \cite{yerokhin:10:helike};\\
$^b$~Yerokhin, Patk\'o\v{s}, and Pachucki 2022 \cite{yerokhin:22:helike};
\end{table}


\appendix

\section{Large-$\bm{Z}$ limit}
\label{sec:app}

To the leading order in the large-$Z$ expansion we can omit all operators containing the electron-electron
radial distance and keep only the electron-nucleus operators containing $r_1$ and $r_2$.
The spatial part of the wave function in the large-$Z$ limit is given by
an (anti-) symmetrized product of two hydrogenic wave functions,
\begin{equation}
	\psi(r_1,r_2) = \frac{1}{\sqrt{2}}\big[\psi_{10}(r_1)\,\psi_{nl}(r_2)\pm\psi_{nl}(r_1)\,\psi_{10}(r_2)\big]\,,
\end{equation}
where the plus sign stands for the singlet and the minus sign, for the triplet states,
and $\psi_{nl}(r)$ are the hydrogenic radial wave functions with the principal quantum number $n$ and
the orbital momentum $l$.
The expectation value of an arbitrary operator $O$ with the triplet-state wave function is
\begin{eqnarray}\label{ap:1}
	\langle O\rangle &=& \frac{1}{2}\langle (1,0),(n,l)|O|(1,0),(n,l)\rangle
\nonumber\\ &&
+ \frac12\langle (n,l),(1,0)|O|(n,l),(1,0)\rangle
\nonumber\\
	&&
-\frac12\langle (n,l),(1,0)|O|(1,0),(n,l)\rangle
\nonumber\\ &&
-\frac12\langle (1,0),(n,l)|O|(n,l),(1,0)\rangle
\,,
\end{eqnarray}
where $|(m,l_1),(n,l_2)\rangle = \psi_{ml_1}(r_1)\,\psi_{nl_2}(r_2)$.

If the operator $O$ is a sum of
one-electron operators $O = O'(r_1) + O'(r_2)$, the first two terms in the right-hand-side of
Eq.~(\ref{ap:1}) are reduced to the sum of two one-electron matrix elements,
$\bra{10}O'\ket{10} + \bra{nl}O'\ket{nl}$. The last two
terms in the right-hand-side of Eq.~(\ref{ap:1}) are of a different form.
It can be shown that for the large-$Z$ limit of the total $m\alpha^7$ correction
such ``mixing'' terms from the first-order operators
cancel identically with the corresponding terms in
the second-order contribution.

For evaluating the large-$Z$ limit of various operators contributing to the $m\alpha^7$ correction,
we make use of the following results for the one-electron matrix elements,
\begin{eqnarray}
	\bra{nl}\frac{1}{r}\ket{nl} &=& \frac{Z}{n^2},\\
	\bra{nl}\frac{1}{r^2}\ket{nl} &=& \frac{Z^2}{n^3(l+\frac12)},\\
	\bra{nl}p^2\ket{nl} &=& 2E_n+\bra{nl}\frac{2Z}{r}\ket{nl} = \frac{Z^2}{n^2},\\
	\bra{nl}4\pi \delta^3(r)\ket{nl} &=& \frac{4Z^3}{n^3}\delta_{l0},\\
	\bra{nl}\vec p\,4\pi \,\delta^3(r)\,\vec p\,\ket{nl} &=& \frac{4Z^5}{3}\bigg(-\frac{1}{n^5}+\frac{1}{n^3}\bigg)\delta_{l1}\,,
\end{eqnarray}
\begin{widetext}
\begin{eqnarray}
	\bra{nl}\frac{1}{r^3}\ket{nl} &=& \frac{4Z^3}{n^3}\bigg(\ln\frac{n}{2Z}-\Psi(n)-\gamma+\frac12-\frac{1}{2n}\bigg)\delta_{l,0}+
	\frac{2Z^3}{n^3} \frac{1-\delta_{l,0}}{l(l+1)(2l+1)},\\
	\bra{nl}\frac{1}{r^4}\ket{nl} &=& \frac{8Z^4}{n^3}\bigg(-\ln\frac{n}{2Z}+\Psi(n)+\gamma-\frac53+\frac{1}{2n}+\frac{1}{6n^2}\bigg)\delta_{l,0}\nonumber\\&&+
	(1-\delta_{l,0})\frac{4Z^4\big(3 n^2-l(1+l)\big)}{(2l - 1)l(2l+1)(l+1)(2l+ 3) n^5},\\
	\bra{nl}p^2\frac{1}{r}\ket{nl} &=& 2E_n\bra{nl}\frac{1}{r}\ket{nl}
	+2Z\bra{nl}\frac{1}{r^2}\ket{nl}=-\frac{Z^3}{n^4}+\frac{2Z^3}{n^3(l+\frac12)},\\
	\bra{nl}\vec p \,\frac{1}{r^2}\vec p\,\ket{nl} &=& Z^4\bigg[\delta_{l0}\bigg(-\frac{2}{3 n^5} + \frac{8}{3n^3}\bigg)
	+\frac{2(1-\delta_{l0})}{(2l-1)(2l+1)(2l+3)}\bigg(\frac{\big(1-4l(l+1)\big)}{n^5}
	+\frac{8}{n^3}\bigg)\bigg]\,.
\end{eqnarray}
\end{widetext}

\end{document}